\documentclass[aps,preprint,superscriptaddress]{revtex4-2}
\usepackage{graphicx}
\usepackage{epsfig}
\usepackage{epstopdf}
\usepackage{bm}
\usepackage{siunitx}
\usepackage{color}
\usepackage[colorlinks,linkcolor=blue, anchorcolor=blue, urlcolor=blue, citecolor=blue]{hyperref}
\usepackage{multirow}
\usepackage{amsmath}
\usepackage{booktabs}
\usepackage{enumerate}
\usepackage{upgreek}

\newcommand\oprod[2]{\ensuremath{|#1\rangle\langle#2|}}
\newcommand\mean[1]{\ensuremath{\langle #1 \rangle}}
\usepackage{array}
\newcommand{\PreserveBackslash}[1]{\let\temp=\\#1\let\\=\temp}
\newcolumntype{C}[1]{\rangle{\PreserveBackslash\centering}p{#1}}
\newcolumntype{R}[1]{\rangle{\PreserveBackslash\raggedleft}p{#1}}
\newcolumntype{L}[1]{\rangle{\PreserveBackslash\raggedright}p{#1}}
\begin{document}

\author{Wei Li}
\thanks{Equal contribution.}
\affiliation{Hefei National Research Center for Physical Sciences at the Microscale and School of Physical Sciences, University of Science and Technology of China, Hefei 230026, China}
\affiliation{Shanghai Research Center for Quantum Science and CAS Center for Excellence in Quantum Information and Quantum Physics, University of Science and Technology of China, Shanghai 201315, China}

\author{Likang Zhang}
\thanks{Equal contribution.}
\affiliation{Hefei National Research Center for Physical Sciences at the Microscale and School of Physical Sciences, University of Science and Technology of China, Hefei 230026, China}
\affiliation{Shanghai Research Center for Quantum Science and CAS Center for Excellence in Quantum Information and Quantum Physics, University of Science and Technology of China, Shanghai 201315, China}

\author{Yichen Lu}
\thanks{Equal contribution.}
\affiliation{Hefei National Research Center for Physical Sciences at the Microscale and School of Physical Sciences, University of Science and Technology of China, Hefei 230026, China}
\affiliation{Shanghai Research Center for Quantum Science and CAS Center for Excellence in Quantum Information and Quantum Physics, University of Science and Technology of China, Shanghai 201315, China}

\author{Zheng-Ping Li}
\affiliation{Hefei National Research Center for Physical Sciences at the Microscale and School of Physical Sciences, University of Science and Technology of China, Hefei 230026, China}
\affiliation{Shanghai Research Center for Quantum Science and CAS Center for Excellence in Quantum Information and Quantum Physics, University of Science and Technology of China, Shanghai 201315, China}
\affiliation{Hefei National Laboratory, University of Science and Technology of China, Hefei 230088, China}

\author{Cong Jiang}
\author{Yang Liu}
\affiliation{Jinan Institute of Quantum Technology, Jinan, Shandong 250101, China}

\author{Jia Huang}
\author{Hao Li}
\author{Zhen Wang}
\affiliation{State Key Laboratory of Functional Materials for Informatics, Shanghai Institute of Microsystem and Information Technology, Chinese Academy of Sciences, Shanghai 200050, China}

\author{Xiang-Bin Wang}
\affiliation{Hefei National Laboratory, University of Science and Technology of China, Hefei 230088, China}
\affiliation{Jinan Institute of Quantum Technology, Jinan, Shandong 250101, China}
\affiliation{State Key Laboratory of Low Dimensional Quantum Physics, Department of Physics, Tsinghua University, Beijing, China}

\author{Qiang Zhang}
\affiliation{Hefei National Research Center for Physical Sciences at the Microscale and School of Physical Sciences, University of Science and Technology of China, Hefei 230026, China}
\affiliation{Shanghai Research Center for Quantum Science and CAS Center for Excellence in Quantum Information and Quantum Physics, University of Science and Technology of China, Shanghai 201315, China}
\affiliation{Hefei National Laboratory, University of Science and Technology of China, Hefei 230088, China}
\affiliation{Jinan Institute of Quantum Technology, Jinan, Shandong 250101, China}

\author{Lixing You}
\affiliation{State Key Laboratory of Functional Materials for Informatics, Shanghai Institute of Microsystem and Information Technology, Chinese Academy of Sciences, Shanghai 200050, China}
\author{Feihu Xu}
\author{Jian-Wei Pan}
\affiliation{Hefei National Research Center for Physical Sciences at the Microscale and School of Physical Sciences, University of Science and Technology of China, Hefei 230026, China}
\affiliation{Shanghai Research Center for Quantum Science and CAS Center for Excellence in Quantum Information and Quantum Physics, University of Science and Technology of China, Shanghai 201315, China}
\affiliation{Hefei National Laboratory, University of Science and Technology of China, Hefei 230088, China}

\title{Twin-field quantum key distribution without phase locking}

\begin{abstract}
	Twin-field quantum key distribution (TF-QKD) has emerged as a promising solution for practical quantum communication over long-haul fiber. However, previous demonstrations on TF-QKD require the phase locking technique to coherently control the twin light fields, inevitably complicating the system with extra fiber channels and peripheral hardware. Here we propose and demonstrate an approach to recover the single-photon interference pattern and realize TF-QKD \emph{without} phase locking. Our approach separates the communication time into reference frames and quantum frames, where the reference frames serve as a flexible scheme for establishing the global phase reference. To do so, we develop a tailored algorithm based on fast Fourier transform to efficiently reconcile the phase reference via data post-processing. We demonstrate no-phase-locking TF-QKD from short to long distances over standard optical fibers. At 50-km standard fiber, we produce a high secret key rate (SKR) of 1.27 Mbit/s, while at 504-km standard fiber, we obtain the repeater-like key rate scaling with a SKR of 34 times higher than the repeaterless secret key capacity. Our work provides a scalable and practical solution to TF-QKD, thus representing an important step towards its wide applications.
\end{abstract}
\maketitle

\textit{Introduction.}---Quantum key distribution (QKD) can provide information-theoretically secure keys among distant parties~\cite{bennett1984quantum} and it has become an indispensable cryptographic primitive in the upcoming quantum era~\cite{xu2020secure,pirandola2020advances}. Due to the loss of photons in their transmission, the point-to-point secret key capacity (SKC$ _\mathrm{0} $) {of a channel} without the quantum repeater scales linearly $ O(\eta) $ with the channel transmission~\cite{pirandola2009direct,takeoka2014fundamental,pirandola2017fundamental}. The twin-field (TF) QKD protocol~\cite{lucamarini_overcoming_2018}, an efficient version of measurement-device-independent QKD~\cite{lo2012measurement}, can greatly enhance the transmission distance by achieving a repeater-like rate-loss scaling of $O(\sqrt{\eta})$ with current available technology. Consequently, TF-QKD has been studied extensively in theory~\cite{ma2018phase,wang2018twin,curty2018simple,cui2019twin} and experiment~\cite{minder2019experimental,liu2019experimental,wang2019beating,fang2020implementation,zhong2019proof,pittaluga2021600,chen2021twin,clivati2022coherent,chen2022quantum,wang_twin-field_2022}. These efforts make the long-haul fiber network within reach. Moreover, its measurement-device-independent advantage can remove trusted nodes from the networks, thus granting a security boost over deployed quantum communication network~\cite{peev2009secoqc,sasaki2011field,chen2021integrated}.

In practice, however, TF-QKD is phase-sensitive~\cite{lucamarini_overcoming_2018}, which normally requires sufficiently long coherence time for two independent laser sources. In previous TF-QKD realizations, such stringent requirement has been fulfilled by phase-locking laser sources using the optical phase-lock loop~\cite{minder2019experimental,wang2019beating,pittaluga2021600,wang_twin-field_2022}, the time-frequency dissemination~\cite{liu2019experimental,chen2021twin,chen2022quantum,clivati2022coherent} or the injection locking~\cite{fang2020implementation} techniques. These approaches require extra servo channels to disseminate the reference light and peripheral hardware to perform the locking, which could potentially hinder the wide deployment of TF-QKD in network settings. Furthermore, the future quantum network may involve users with free-space link~\cite{vallone2015experimental,cao2020long} or with integrated photonic chip~\cite{ma2016silicon,sibson2017chip,wei2020high}. However, the servo channel is hard to establish in the free-space link, and the phase-locking components are challenging to be integrated on chip.

In this work, we propose and demonstrate an approach to realize TF-QKD \emph{without} phase locking. We alternate the communication period into quantum frame (Q-frame) and reference frame (R-frame), and use the R-frame to provide a phase reference for the Q-frame by reconciling the signals via data post-processing. We develop an algorithm based on the fast Fourier transform (FFT) to efficiently track the frequency and the phase fluctuation. By doing so, we are able to recover the interference pattern with 259 photon detections in a duration of 7 $ \upmu $s, yielding an interference error rate (ER) of 2.32\%, or an ER of 3.74\% with 159 photon detections in a duration of 11 $ \upmu $s. Moreover, we derive an analytical model to study the sources of phase fluctuation and provide a guideline to optimize the experimental parameters in the aspect of the ratio of the Q-frame to the R-frame. To test our approach, we build a TF-QKD setup without any phase locking (or phase compensation) at the transmitters (or receiver), and demonstrate TF-QKD from 50- to 504-km of standard fiber channels. A secret key rate (SKR) of 2.05 bit/s is generated over 504-km standard fiber (96.8 dB loss or equivalent to 605-km ultra low loss fiber) in the finite-size regime, which is 34 times higher than SKC$ _\mathrm{0} $. At 50-km standard fiber, we are able to produce 1.27 Mbit/s secret keys.

\textit{No-phase-locking scheme.}---The phase difference of two light fields $ \Phi(t) $ evolves as,
\begin{equation}\label{eq:phi}
\Phi(t) = 2\pi \nu_0 t+ \phi_0 + \Delta \phi(t),
\end{equation}
where $\nu_0$, $\phi_0$ and $\Delta \phi(t)$ denote the beat-note frequency, the initial phase and the phase fluctuation respectively. {Both $\nu_0$ and $\phi_0$ are constant and can be estimated via the data post-processing,} whereas $\Delta \phi(t)$ includes the intrinsic phase noise of the laser sources and the fluctuation introduced in the channel transmission.
\begin{figure}[!htb]
	\centering
	\includegraphics[width=\linewidth]{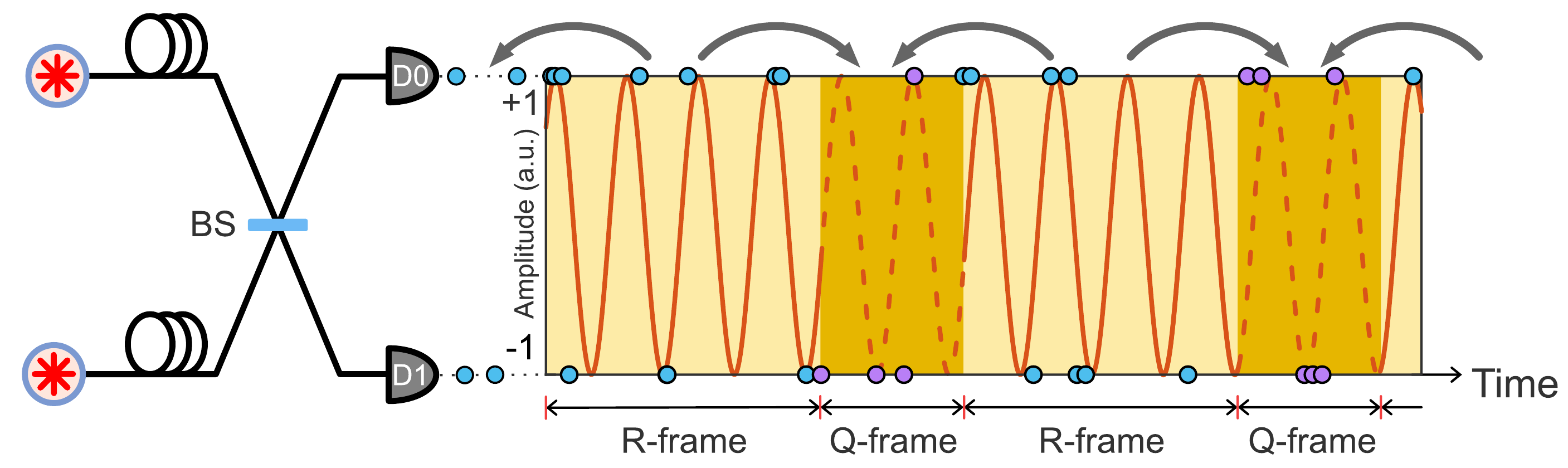}
	\caption{No-phase-locking scheme. The interference pattern is evaluated frequently in the R-frame, providing a phase reference for the Q-frame. The detection events of D$_0$ (D$_1$) are mapped to the amplitude of +1 (-1) and the detection probability is related to the interference pattern. Each R-frame is used twice for the neighboring Q-frames.}
	\label{fig:scheme}
\end{figure}

To perform the parameter estimation, we propose to supplement the quantum pulses (Q-frame) with strong R-frame. The R-frame is used to reconcile the phase in the Q-frame through post-processing the photon detection events. As shown in Fig.~\ref{fig:scheme}, the detection probabilities of two detectors are correlated with the interference pattern. In light of this, we develop an FFT-based algorithm to reconstruct the frequency spectrum (see Supplementary Section II~\cite{supplemental}). From the frequency spectrum, we choose the component with the largest amplitude within a certain frequency range. {Then, $\nu_0$ is the frequency of the component and $\phi_0$ can be obtained from the angle of the complex amplitude of the component}. Here to reduce the estimation error, we propose to extend the detection series with padding zeros. This can greatly narrow the distribution of the frequency estimation and decrease the frequency deviation, thus enhancing the frequency estimation precision (see Supplementary Fig. 1). Moreover, we duplex the detection events to use them more efficiently, i.e., each R-frame is used twice by the two neighbour Q-frame. This doubles the available photon events and reduces the error rate when the photon count rate is low.

According to the estimated $\hat{\nu}_0$ and $\hat{\phi}_0$, the ER can be further evaluated. {When single photon events arrival time $ t $ satisfy (a) $\cos{\left(\hat{\Phi}(t)\right)}\geq\cos{(2\pi/M)}$ or (b) $\cos{\left(\hat{\Phi}(t)\right)}\leq\cos{((M-2)\pi/M)}$, a valid event is counted, where $ \hat{\Phi}(t) = 2\pi \hat{\nu}_0 t+ \hat{\phi}_0 $ is the estimated phase difference and $ M $ is the number of discrete phase slices in the post-selection.} If condition (a) ((b)) is satisfied and D$_0$ (D$_1$) clicks, it is counted as a correct event. Otherwise, if condition (a) ((b)) is satisfied and D$_1$ (D$_0$) clicks, it is counted as an error event.

{The ER originates from the difference $\Phi(t)-\hat{\Phi}(t)=2\pi(\nu_0-\hat{\nu}_0)t+(\phi_0-\hat{\phi}_0)+\Delta\phi(t)$.} The third term (mainly contributed by the phase noise of laser source and the fiber length fluctuation) is a random noise, its contribution can be analyzed in theory and characterized in experiment. To do so, we derive an analytical model to study the phase noise/fluctuation and provide a guideline to optimize the experimental parameters (see Supplementary Section I). {The phase fluctuation can be analyzed in time domain~\cite{lucamarini_overcoming_2018,liu2019experimental,fang2020implementation} or frequency domain~\cite{clivati2022coherent}. Here, we use the frequency-domain power spectral density (PSD) to characterize the random noise.} By doing so, we are able to calculate the fluctuation in variable interval $ \tau $ and relate it to the ER. 
Furthermore, the phase noise of the laser sources is also taken into consideration in our model. This was often ignored in the previous analysis for the phase-locked TF-QKD schemes~\cite{lucamarini_overcoming_2018,liu2019experimental,fang2020implementation}. Nonetheless, as the linewidth of the laser source increases, it will dominate the phase fluctuation. Even for the phase-locked laser sources, their phase noise replicate the one of the reference source at best, which can not be ignored. By including the phase noise, our model can properly analyze the linewidth requirement for the laser sources for both phase-locking and no-phase-locking TF-QKD.

\begin{figure*}[!htb]
	\centering
	\includegraphics[width=\linewidth]{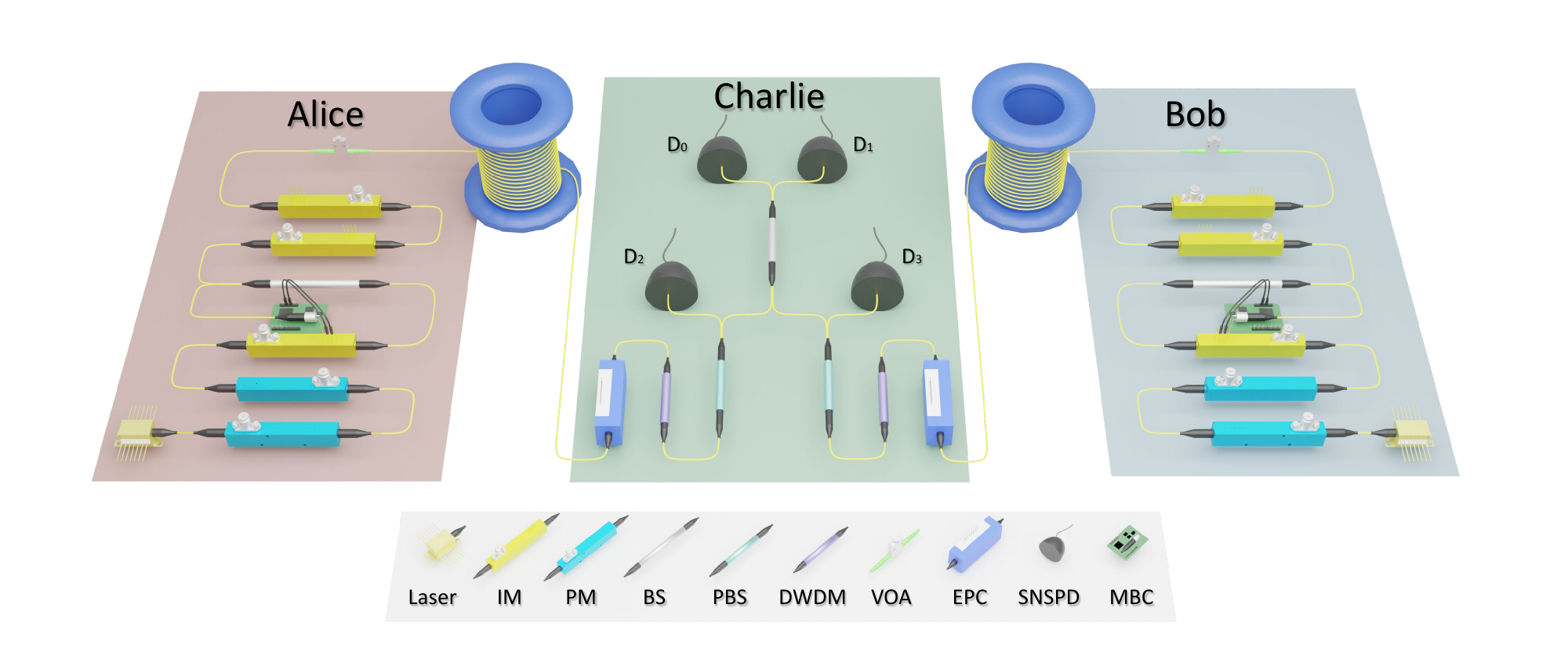}
	\caption{Experimental setup. Alice (Bob) send their phase- and intensity-modulated states to Charlie to perform single-photon measurements. D$ _{0 (1)}$ are used to generate secure keys, while D$ _{2 (3)}$ are used for delay and polarization compensation. PM, phase modulator; IM, intensity modulator; BS, beamsplitter; VOA, variable optical attenuator; EPC, electrical polarization controller; DWDM, dense wavelength-division multiplexer; PBS, polarization beamsplitter; SNSPD, superconducting nanowire single photon detector; MBC, modulator bias controller.}
	\label{fig:setup}
\end{figure*}

\textit{Setup.}---To implement the no-phase-locking TF-QKD scheme, we build an experimental setup as shown in Fig.~\ref{fig:setup}. Alice and Bob transmit their quantum light to Charlie's measurement site via symmetric quantum channels, constituted by standard telecom fiber spools (G.652). No servo link is used for the dissemination of a third phase reference laser. Each user holds a commercial external cavity laser diode (RIO PLANEX) with a Lorentzian linewidth of 5 kHz and the wavelength is set at 1550.12 nm, but with a slight frequency mismatch of about 100 MHz. The frequency drifts slowly to an extent of 30 MHz over one day (see Supplementary Section VI), which can be tracked by the frequency estimation algorithm.

The continuous light is encoded into two frames by three cascaded intensity modulators. Due to the fact that two lasers are heterodyned, no modulation in intensity and phase is required for the R-frame to resolve the phase ambiguity as in ref.~\cite{liu2019experimental,fang2020implementation}. The Q-frame is used to generate quantum signals following the 3-intensity sending-or-not-sending (SNS) TF-QKD protocol~\cite{yu2019sending} (see Supplementary Section VII for full descriptions) with a clock rate of 1.25 GHz. Two cascaded phase modulators are used for 16-level random phase modulation covering 2$ \pi $ range. One intensity modulator creates the two frames and modulates the weak decoy intensity {(if the intensity contrast is over the capability of one modulator, the weak decoy is modulated by another modulator instead, see Supplementary Table III)}, and the other two shape the pulses in Q-frame and extinct vacuum pulses jointly. Such configuration allow us to apply bias control on the first intensity modulator to stabilize the signal and decoy intensities.

Light from Alice and Bob interfere at Charlie's 50:50 beam splitter with the same polarization as the output of the polarization beamsplitter is polarization-maintaining. This transforms the polarization variation into intensity variation. We use two electrical polarization controllers to rotate the polarization at 5 Hz so as to keep the photon rates at the other port of the polarization beamsplitter constant (5\% of total photon rate). Moreover, the photons detected by D$ _2 $ (D$ _3 $) are accumulated to track the delay drift of the fiber every 5 s. The transition edge from the Q-frame to the R-frame is used as the time mark. These are sufficient to compensate the polarization and delay drift caused by the indoor fiber spools.

The interference output is detected by two superconducting nanowire single photon detectors (SNSPD). We use two types of SNSPDs, SNSPD\#1 and \#2, to perform the measurements (see Supplementary Section VIII for the characterization). The detected photon events are registered by a time tagging unit and then processed by a computer. The frequency $ \hat{\nu}_0 $ and initial phase $ \hat{\phi}_0 $ of the beat note can be both estimated by the FFT-based algorithm. The valid arriving time $ t $ in X basis should satisfy:
\begin{equation}
\left|\cos{\left( 2\pi \hat{\nu}_0 t+ \hat{\phi}_0+(\phi_{a}-\phi_{b})\right)}\right| \geq\cos{(\pi/16)},
\end{equation}
where $ \phi_{a} \ (\phi_{b})$ is phase modulated by the users and announced publicly in the post-processing. We also use actively odd-parity pairing method~\cite{xu2020sending,jiang2021composable} in the error rejection through two-way classical communication, significantly reducing the bit-flip error rate.

\textit{Results.}---We firstly analyze the system performance with the no-phase-locking scheme quantitatively. The phase noise PSD of the laser has a -20 dB/decade slope, contributed by the white frequency noise. The phase noise PSD introduced by the fiber spool is also measured (see Supplementary Section III and IV), which is mainly distributed below the frequency 100 kHz and increase slightly with the fiber length. With these two results, the phase fluctuation in the Q-frame can be calculated accordingly. The fluctuation is converted to the ER and averaged over the duration of the Q-frame.

\begin{figure}[htbp]
	\centering
	\includegraphics[width=\linewidth]{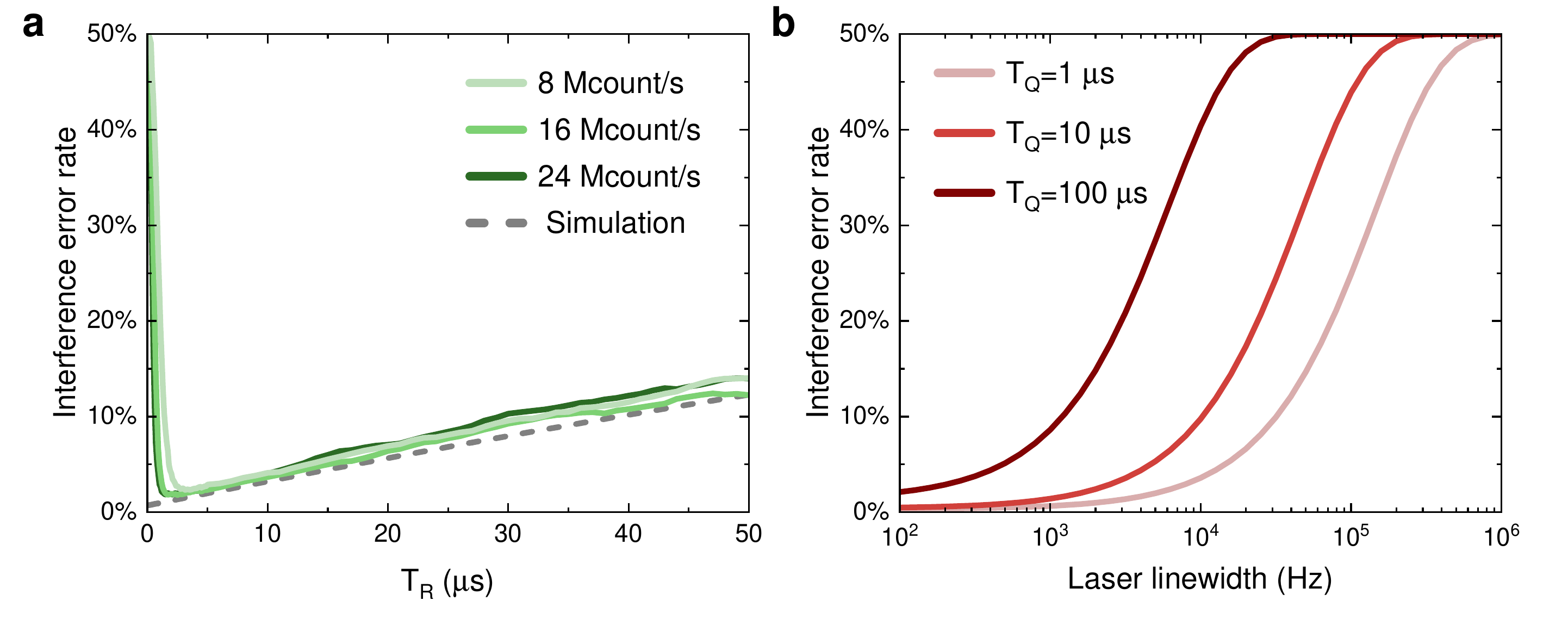}
	\caption{\textbf{a,} {Measured and simulated interference error rate as a function of $T_\mathrm{R}$ from $0.1~\upmu $s to $50~\upmu $s with different count rates per detector and $T_\mathrm{Q}=1~\upmu$s. The simulated results exclude the estimation error and we used the phase noise result of 380-km fiber spools and the lasers linewidth are 5 kHz with a Lorentzian lineshape in the simulation and measurement.} \textbf{b,} Interference error rate as functions of the laser linewidth and $T_\mathrm{R}=5~\upmu$s {in which the estimation error is also excluded and the phase noise result of 380-km fiber spools is used as the channel noise.}}
	\label{fig:ER_all}
\end{figure}

To characterize the ER with different $T_\mathrm{Q}$ and $T_\mathrm{R}$, we use the same setup as in Fig.~\ref{fig:setup}, except that no modulation is applied. {We plot measured ER as a function of $T_\mathrm{R}$ from $0.1~\upmu $s to $50~\upmu $s at a fixed $T_\mathrm{Q}$ of $1~\upmu $s with different photon count rates in Fig.~\ref{fig:ER_all}a.} A higher count rate enables accurate frequency estimation at smaller $T_\mathrm{R}$, thus lowering the minimum ER. {The simulation result excluding the estimation error is also plotted, assuming that $\hat{\nu}_0=\nu_0$ and $\hat{\phi}_0=\phi_0+\overline{\Delta\phi(t)}$ (see Supplementary Section I).} Based on these results and the constraint of the experimental system, we choose $T_\mathrm{R}$ and $T_\mathrm{Q}$ to be 4.9152 and 1.6384 $ \upmu $s respectively. The ER contributes to the quantum bit error rate (QBER) of phase (X) basis in SNS-TF-QKD protocol. As another important implication of our theoretical model, we can determine the linewidth requirement for different selected $T_\mathrm{Q}$ and plot the simulated results in Fig.~\ref{fig:ER_all}b. As an example, to achieve a maximum channel loss of 66 dB in our system, the laser linewidth should be narrower than 35.5 kHz when $T_\mathrm{Q}=1~\upmu $s, resulting in an ER of 11\%.

\begin{figure}[htbp]
	\small
	\centering
	\includegraphics[width=\linewidth]{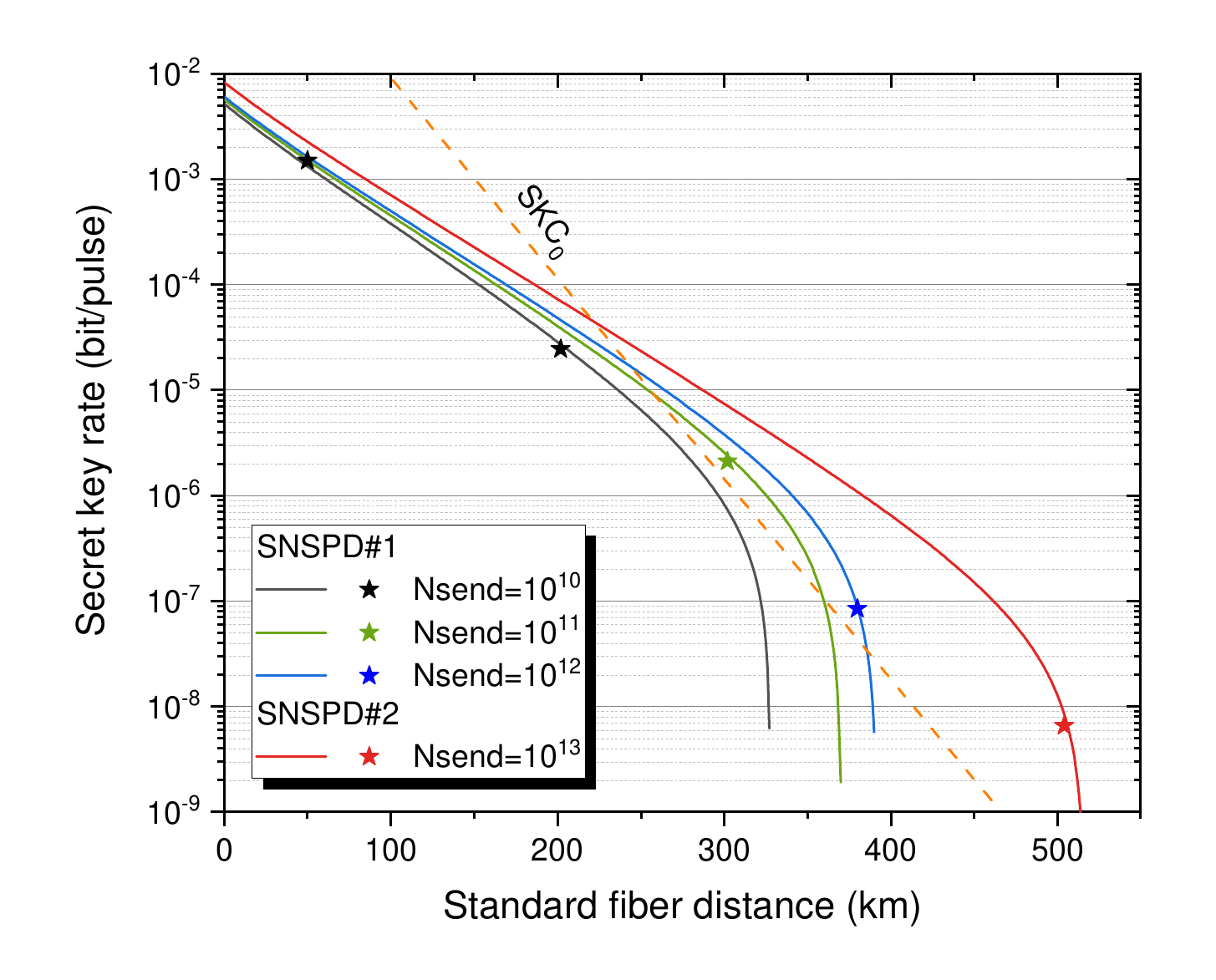}
	\caption{SKRs at different standard fiber distances. The solid line is the simulated SKRs with different finite sizes and two sets of experimental parameters. The fiber loss coefficient is 0.19 dB/km. The solid stars denote the experimental results. The orange dashed line denotes SKC$ _\mathrm{0}$\cite{pirandola2017fundamental}.}
	\label{fig:rate}
\end{figure}

With the performance analysis and parameter optimization, we perform TF-QKD experiments from 50 to 504-km standard fiber spools with different detectors and finite sizes (Supplementary Section VIII). The channel is symmetrical and the total channel loss amounts to 9.6, 38.4, 56.8, 72.1 and 96.8 dB respectively. The results are presented in Fig.~\ref{fig:rate}. With SNSPD\#1, we measure the dynamic range to be 44 dB when the count rate of the R-frame is 24 Mcount/s and the duty cycle (i.e., the ratio of Q-frame) is 1/4. That is to say the dynamic dark count rate~\cite{burenkov_investigations_2013,chen2015dark} is about 1000 count/s, which is the dominant noise. To increase the signal-to-noise ratio, we apply a gating window of 200 ps, resulting in a dark count probability of $ 2\times10^{-7} $. With less than one hour of continuous run, we achieve a finite-size SKR of $ 1.25\times 10^{-7} $ bit/pulse, or 39 bit/s at 380-km standard fiber. At the short end, we increase the duty cycle {of the Q-frame to 3/4 to enhance the per-second key rate.} And it enable a secret key rate of 1.27 Mbit/s at 50-km standard fiber channel with a small number of sending pulses of $ 10^{10} $.

To generate positive key rates at longer distance, we replace the detectors with SNSPD\#2, which has a dynamic range over 58 dB. Furthermore, we keep the detected count rate in R-frame as low as possible (8 Mcounts/s per detector) and use a narrower gating window of 100 ps. With these upgrades, the scattering noise in fiber becomes dominant and results in a noise probability of $ 1.6\times10^{-8} $. This is an order of magnitude of improvement over SNSPD\#1, allowing us to achieve a SKR of $ 6.56\times 10^{-9} $ bit/pulse or 2.05 bit/s at 504-km standard optical fiber.

\textit{Discussion.}---In summary, we have proposed and demonstrated the no-phase-locking scheme for TF-QKD. Our scheme does not only remove the service channels for the dissemination of the reference light, but also does not needs active phase compensation setup at the measurement site. Such features greatly simplify the setup to match the one of measurement-device-independent QKD systems~\cite{liu2013experimental,comandar_quantum_2016}, with the ability to establish global phase reference nonetheless. We also show that commercial kilohertz linewidth semiconductor lasers are sufficient to perform TF-QKD with our scheme. {Despite the simplification of setup with our scheme, the phase-sensitive ER is 2.69\% for a channel distance of 504 km (Supplementary Table I) which is comparable to the state-of-art TF-QKD experiment with phase locking~\cite{pittaluga2021600,chen2022quantum,wang_twin-field_2022}}. To achieve further transmission distance with our scheme, {one could increase the clock rate and develop more advanced algorithm to lower the required counts in the recovery of the carrier, which helps to reduce the influence of scattering noise.} Overall, we believe our scheme provides a practical solution to TF-QKD network and the phase recovery algorithm is applicable to other phase-sensitive applications~\cite{duan_long-distance_2001,humphreys_deterministic_2018}.

{While preparing the manuscript, we notice two related works which demonstrate different approaches~\cite{zhu_experimental_2023,PhysRevLett.130.250801} to solve the phase-locking issue. Comparing to Ref.~\cite{zhu_experimental_2023}, the setup has similarities, but the protocols are different. Our work uses the twin-field protocol, but Ref.~\cite{zhu_experimental_2023} employs the so-called mode-pairing protocol. The mode-pairing protocol is a post-pairing approach by postselecting time slots within the coherence time of the laser source. The post-pairing distance for time slots has to increase with the transmission loss, which leads to a large interference error due to the limited coherence time of the laser. As a result, our system endures higher transmission loss (96.8 dB) comparing to Ref.~\cite{zhu_experimental_2023} (66 dB). This can be compensated using ultra-stable lasers which have a much longer coherence time~\cite{PhysRevLett.130.250801}, but adding more complexity and cost. In contrast, we demonstrate an efficient method to reconcile the phase reference in TF-QKD protocol. Our approach has no stringent requirements for the laser source, but it requires the estimation of full phase information which may need further software processing and sufficient photon counts.}

\textit{Acknowledgement.}--- This work was supported by National Natural Science Foundation of China (Grant No. 62031024, No. 12204467), Innovation Program for Quantum Science and Technology (2021ZD0300300), Anhui Initiative in Quantum Information Technologies, Shanghai Municipal Science and Technology Major Project (Grant No. 2019SHZDZX01), Shanghai Science and Technology Development Funds (22JC1402900) and Shanghai Academic/Technology Research Leader (21XD1403800).  Y.L. acknowledges support from the Taishan Scholar Program of Shandong Province. F. Xu acknowledges the support from the Tencent Foundation.

\clearpage
\setcounter{equation}{0}
\setcounter{table}{0}
\setcounter{figure}{0}
\renewcommand{\thefigure}{S\arabic{figure}}
\renewcommand{\thetable}{S\arabic{table}}
\begin{center}
	\textbf{{\large Supplementary for ``Twin-field quantum key distribution without phase locking"}}
	\vspace{0.5cm}
\end{center}
\section{Interference error rate calculation}
Here we present the theoretical model for the calculation of the interference error rate introduced by the phase fluctuation. The emitted laser electric field undergoing phase fluctuations at Alice's (Bob's) side can be written as
{\begin{equation}
	\label{P2p}
	\begin{aligned}
	E_{\mathrm{A(B)}}(t)&=E_0 \exp{\left\{i\Phi_\mathrm{A(B)}(t)\right\}},
	\\
	\Phi_\mathrm{A(B)}(t)& = \int_{0}^{t} 2\pi\nu_\mathrm{A(B)}(t^{\prime}) \mathrm{d} t^{\prime} + \phi_\mathrm{0A(B)},
	\end{aligned}
	\end{equation}
	where $\Phi_\mathrm{A(B)}(t)$, $ \nu_\mathrm{A(B)}(t) $ and $\phi_\mathrm{0A(B)}$ are the phase, the frequency and the initial phase of the laser, $\nu_\mathrm{A(B)}(t)$ is a stochastic process representing the frequency noise of the laser.}


After transmitting through the fiber, the fluctuation of the fiber length and refractive index changes the optical distance, and therefore changes the phase of the light field. Note that the light right before interference at time $t$ is emitted at time $t-L_\mathrm{A(B)}(t)/c$, where $L_\mathrm{A(B)}(t)$ is the optical distance through which the photon arriving at time $t$ passes and $c$ is the speed of light in vacuum. We assume $L_\mathrm{A(B)}(t)$ to be a Gaussian stationary process. The phase change of the laser is the integral of frequency within the time $\tau$:
\begin{equation}
\label{model}
\begin{aligned}
&\Phi_\mathrm{A(B)}(t+\tau)-\Phi_\mathrm{A(B)}(t)
\\
=&\int_{t-L_\mathrm{A(B)}(t)/c}^{t+\tau-L_\mathrm{A(B)}(t+\tau)/c} 2\pi\nu_\mathrm{A(B)}(t^{\prime}) \mathrm{d} t^{\prime}
\\
=&\int_{t-L_\mathrm{A(B)}(t)/c}^{t+\tau-L_\mathrm{A(B)}(t+\tau)/c} 2\pi\delta\nu_\mathrm{A(B)}(t^{\prime}) \mathrm{d} t^{\prime} +2\pi\nu_\mathrm{0A(B)}\cdot(\tau-\frac{L_\mathrm{A(B)}(t+\tau)-L_\mathrm{A(B)}(t)}{c}),
\end{aligned}
\end{equation}
where $ \delta\nu_\mathrm{A(B)}(t)=\nu_\mathrm{A(B)}(t)-\left<\nu_\mathrm{A(B)}\right> $. {We assume $\delta\nu_\mathrm{A(B)}(t)$ is a zero-mean wide-sense stationary process. $\nu_\mathrm{0A(B)}=\left<\nu_\mathrm{A(B)}\right>$ and $\left<\nu_\mathrm{A(B)}\right> $ is the ensemble average of $\nu_\mathrm{A(B)}$. It is the result of treating the frequency modulation noise $\delta\nu_\mathrm{A(B)}(t)$ as the source of the phase noise. And the phase noise can be related to the frequency noise in the following form:
	\begin{equation}
	2\pi\nu_\mathrm{0A(B)} t+\phi_\mathrm{A(B)}(t)
	= \int_{0}^{t} 2\pi\nu_\mathrm{A(B)}(t^{\prime}) \mathrm{d} t^{\prime} + \phi_\mathrm{0A(B)}.
	\end{equation}
}By substituting Equation~\ref{P2p} and \ref{model} into above equation, we can get phase fluctuation at one side:
\begin{equation}
\begin{aligned}
\Delta\phi_\mathrm{A(B)}(\tau)=&\phi_\mathrm{A(B)}(t+\tau)-\phi_\mathrm{A(B)}(t)
\\
=&\int_{t-L_\mathrm{A(B)}(t)/c}^{t+\tau-L_\mathrm{A(B)}(t+\tau)/c} 2\pi \delta\nu_\mathrm{A(B)}(t^\prime) d t^\prime-2\pi\nu_\mathrm{0A(B)}\frac{L_\mathrm{A(B)}(t+\tau)-L_\mathrm{A(B)}(t)}{c}.
\end{aligned}
\end{equation}

%

By using the Wiener–Khinchin theorem and the Gaussian stationary assumptions, its second order moment has the following form:
\begin{equation}
\label{eq:phase_error}
\begin{aligned}
&\left<\Delta\phi_\mathrm{A(B)}^2(\tau)\right>
\\
=&4\pi^2\tau^2 \int_{0}^{\infty} S_{\delta \nu_\mathrm{A(B)}}(f)\operatorname{sinc}^2(\pi f \tau) \mathrm{d}f
\\
+&4\pi^2\nu_\mathrm{0A(B)}^2 \sigma_{L_\mathrm{A(B)}}^2(\tau)
\\
+&2\int_{0}^{\infty}
S_{\delta \nu_\mathrm{A(B)}}(f) \left[1-\exp{\left(-2\pi^2\sigma_{L_\mathrm{A(B)}}^2(\tau)f^{2}\right)}\right]
\frac{\cos{\left(2\pi f \tau\right)}}{f^{2}}
\mathrm{d}f,
\end{aligned}
\end{equation}
{where $S_{\delta\nu_\mathrm{A(B)}}(f)$ is the power spectral density (PSD) of the laser frequency fluctuation $\delta\nu_\mathrm{A(B)}$, and $\sigma_{L_\mathrm{A(B)}}^2(\tau) =\left<\left(\frac{L_\mathrm{A(B)}(\tau)-L_\mathrm{A(B)}(0)}{c}\right)^2\right>_{L_\mathrm{A(B)}}$. $S_{\delta\nu_\mathrm{A(B)}}(f)$ and $S_{\phi}(f)$ can be obtained from the PSD measurement of laser and fiber phase noise accordingly.}

In the following, we omit the subscript to represent the difference of two sides. Since the noise of Alice's side and Bob's side are identical and independent, so the variance of phase fluctuation at time $t+\tau$ can be calculated:
\begin{equation}
\left<\Delta\phi^2(\tau)\right>
:=\left<\left(\phi(t+\tau)-\phi(t)\right)^2\right>
=\left<\Delta\phi_\mathrm{A}^2(\tau)\right> + \left<\Delta\phi_\mathrm{B}^2(\tau)\right>
\end{equation}

{In the experiment, we need the R-frame with time interval $\mathcal{R}=[-T_\mathrm{R}-T_\mathrm{Q}/2, -T_\mathrm{Q}/2]\cup[T_\mathrm{Q}/2, T_\mathrm{Q}/2+T_\mathrm{R}]$ to estimate $\phi$ as $\hat{\phi}$, $\nu$ as $\hat{\nu}$  and predict the relative phase in the Q-frame $\mathcal{Q}=[-T_\mathrm{Q}/2,T_\mathrm{Q}/2]$. The estimation process can be modeled as the following least square optimization problem assuming the estimate of frequency difference is accurate $\hat{\nu}=\nu_0=\nu_\mathrm{0A}-\nu_\mathrm{0B}$ :}
\begin{equation}
\begin{aligned}
&\frac{1}{2T_\mathrm{R}}\int_\mathcal{R} \left[\hat{\phi}-\phi(t)\right]^2 \mathrm{d}t
\\
=&\frac{1}{2T_\mathrm{R}}\int_\mathcal{R} \left[\left(\phi(t)-\frac{1}{2T_\mathrm{R}}\int_\mathcal{R} \phi(t^\prime) \mathrm{d} t^\prime\right)+\left(\frac{1}{2T_\mathrm{R}}\int_\mathcal{R} \phi(t^\prime) \mathrm{d} t^\prime-\hat{\phi}\right)\right]^2 \mathrm{d}t
\\
=&\overline{\Delta \phi^2}+\left(\bar{\phi}-\hat{\phi}\right)^2
\end{aligned}
\end{equation}
where $\overline{\Delta \phi^2}=\frac{1}{2T_\mathrm{R}}\int_\mathcal{R} \left(\phi(t^\prime)-\bar{\phi}\right)^2 \mathrm{d} t^\prime$ is the time average of variance over R-frame. Note that this is different from the ensemble average of variance $\left<\Delta \phi^2\right>$. {To exclude the estimation error caused by the estimation algorithm and limited photon counts in the simulation, we need to minimize the above formula.} $\hat{\phi}$ takes the following value,
\begin{equation}
\hat{\phi}=\bar{\phi}=\frac{1}{2T_\mathrm{R}}\int_\mathcal{R} \phi(t^\prime) \mathrm{d} t^\prime
\end{equation}

The phase fluctuation in the quantum frame (at time $t\in\mathcal{Q}$) is derived as follow:
\begin{equation}
\label{eq:phiR}
\begin{aligned}
\left< \left( \phi(t)-\hat{\phi}\right)^2\right>
=&\frac{1}{(2T_\mathrm{R})}
\int_0^T  \left[ \left< \Delta\phi^2(t^\prime+T_\mathrm{Q}/2+t)\right>+\left< \Delta\phi^2(t^\prime+T_\mathrm{Q}/2-t)\right>\right]\mathrm{d}t^\prime
\\
-&\frac{1}{(2T_\mathrm{R})^2}
\int_0^T  \left[  \left< \Delta\phi^2(T_\mathrm{Q}+t^\prime)\right>+\left< \Delta\phi^2(T_\mathrm{Q}+2T_\mathrm{R}-t^\prime)\right>+2\left< \Delta\phi^2(T_\mathrm{R}-t^\prime)\right>  \right]  t^\prime \mathrm{d}t^\prime
\end{aligned}
\end{equation}

To this end, we have obtained the phase fluctuation after the estimation of the phase reference. And we can prove that by using neighboring R-frames, better estimation can be obtained. The next step is to transform the phase fluctuation to the interference error rate. The instantaneous interference error rate at time $t\in\mathcal{Q}$ is
\begin{equation}
\begin{aligned}
ER_{ins}(t)=\frac{M}{4\pi}\int_{\hat{\phi}=-\frac{2\pi}{M}}^{\frac{2\pi}{M}}
\int_{\phi=-\infty}^{\infty}
\mathrm{sin}^2\left(\frac{\phi}{2}\right)\cdot \rho(\hat{\phi}-\phi;t)
\mathrm{d} \phi
\mathrm{d} \hat{\phi}
\end{aligned}
\end{equation}
where $\rho(\hat{\phi}-\phi;t)$ is the Gaussian probability distribution function of the phase fluctuation and $ M $ is the number of discrete phase slices in the post-selection.
\begin{equation}
\rho(\hat{\phi}-\phi;t)=\dfrac{1}{\sqrt{2\pi\left< \left( \phi(t)-\hat{\phi}\right)^2\right>}}\cdot \exp{\left[-\dfrac{ (\hat{\phi}-\phi)^2}{2\left< \left( \phi(t)-\hat{\phi}\right)^2\right>}\right]}
\end{equation}

Finally, we can obtain the average interference error rate over the Q-frame $\mathcal{Q}=[-T_\mathrm{Q}/2,T_\mathrm{Q}/2]$:
\begin{equation}
ER(\mathcal{Q})=\frac{1}{T_\mathrm{Q}}\int_\mathcal{Q}
\frac{M}{4\pi}\int_{\hat{\phi}=-\frac{2\pi}{M}}^{\frac{2\pi}{M}}
\int_{\phi=-\infty}^{\infty}
\mathrm{sin}^2\left(\frac{\phi}{2}\right)\cdot \rho(\hat{\phi}-\phi;t)
\mathrm{d} \phi
\mathrm{d} \hat{\phi}
\mathrm{d} t
\end{equation}

\section{FFT-based algorithm}
In this experiment, Alice and Bob send bright continuous-wave light to Charlie during the R-frame, who will then broadcast the detection result including which detector clicks and the arrival time. We estimate the frequency and phase of the interference pattern in the post-processing procedure based on these detection results. The estimation problem can be modeled as following. Supposing that the frequency difference of the two independent laser source of Alice and Bob is $\nu$, and the initial phase difference at time $t=t_0$ is $\phi_{0}$, The detection probabilities of the two detector are:
\begin{equation}
\left\{
\begin{aligned}
&P_{0}(t)=\dfrac{1+\cos{\Phi(t)}}{2}\\
&P_{1}(t)=\dfrac{1-\cos{\Phi(t)}}{2}\\
\end{aligned}
\right.
\quad,\quad\\
\end{equation}
where $ \Phi(t)=2\pi \nu (t-t_0)+\phi_{0} $. To recover $ \Phi(t) $ using the detections of two detectors, we defined $P(t)=P_{0}(t)-P_{1}(t)=\cos{\Phi(t)}$. The value of $ i $-th detection event $ x_i $ is assigned to +1 (-1) if D$ _0 $ (D$ _1 $) clicks. And the value of the $ j $-th bin $ X_j=\sum x  $ for all the events in the bin. And we perform the FFT algorithm on the sparse signal $X_j$. From the spectrum, we select the frequency component with the largest amplitude as the estimate of $\hat{\nu}$ and its phase in the spectrum as the estimate of $\hat{\phi}_0$. We set the bin width to 100 ps in the experiment.

Due to the sparsity of $X_j$, the signal to noise ratio is often limited. And the frequency resolution of the spectrum is limited by the time duration of the sample. As a consequence, when there are enough detection events within the R-frame, the precision of the estimated frequency is mainly affected by the resolution of the spectrum. In this case, further intensifying the reference light to increase effective number of detection is no longer beneficial. Another possible way to reduce the frequency estimation error is to combine a number of neighboring R-frames for a single frequency estimation. However, this method is not always useful as the estimation error of $\Phi(t)$ will accumulate with time under a fixed estimation error of $\hat{\nu}$, thus increasing the interference error rate with more R-frame used.

\begin{figure}[htbp]
	\centering
	\includegraphics[width=\linewidth]{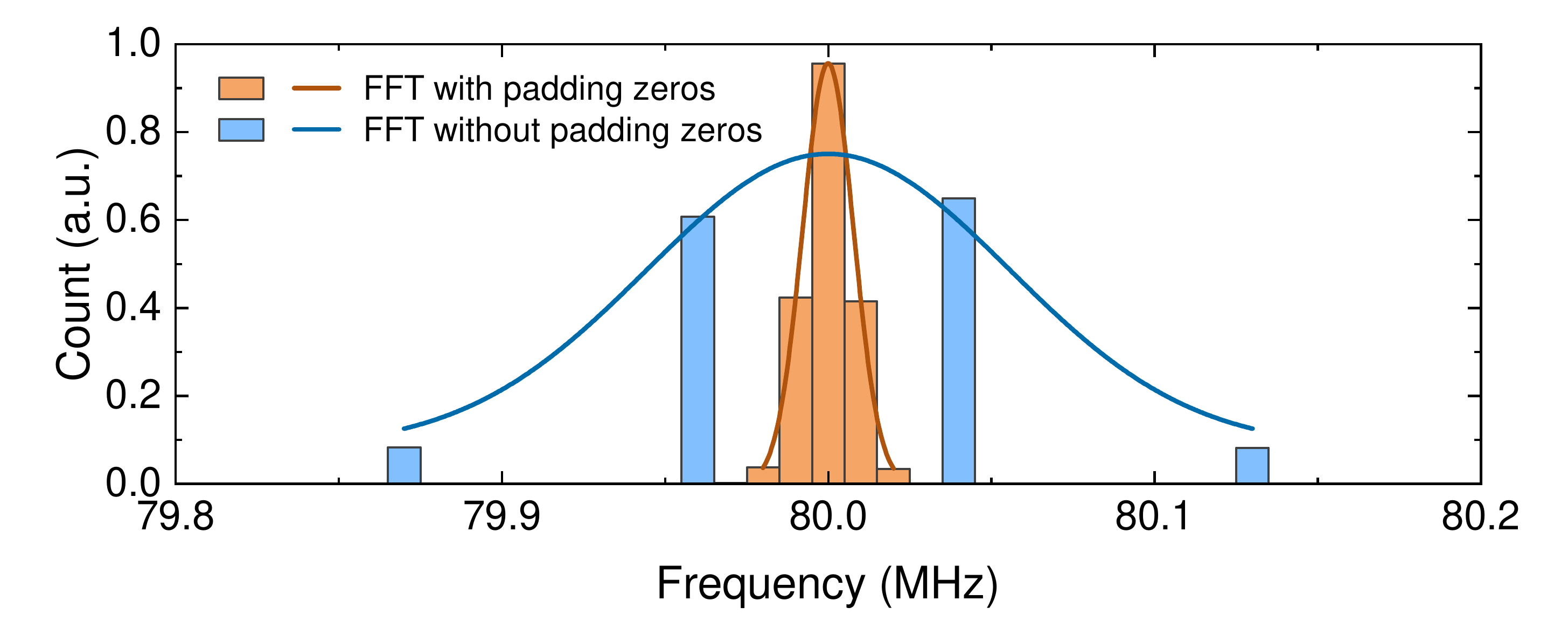}
	\caption{The histogram of frequency estimation results using FFT algorithm with/without padding zeros. The results are based on a beat frequency of 80 MHz. The solid lines represent the corresponding Gaussian fit of the two histograms.}
	\label{fig:FFT_comparison}
\end{figure}

We enhance the spectral resolution by padding $X_n$ with trailing zeros to a suitable length. Fig.~\ref{fig:FFT_comparison} shows the comparison of the frequency estimation results using the FFT algorithm \emph{with} and \emph{without} padding zeros. By improving the resolution from about 0.1 MHz to 0.01 MHz, the full width at half maximum of the frequency distribution decreases from 0.13 MHz to 0.02 MHz. It is clear that padding zeros greatly narrows the distribution of the frequency estimation and decreases frequency deviation. This in turn verifies that the frequency estimation precision is mainly limited by the frequency resolution of the algorithm instead of our system. Besides, in our experimental implementation, we utilize two neighboring R-frames to obtain one estimate of the differential phase of the R-frame. This doubles the available photon events for the estimation while contributing little to the error accumulation effect, which reduce the error rate with relatively low count rate at longer distance. {Moreover, even excluding estimation error (i.e. $\hat{\nu}=\nu_0$, $\hat{\phi}=\bar{\phi}=\frac{1}{2T_\mathrm{R}}\int_\mathcal{R} \phi(t^\prime) \mathrm{d} t^\prime$), the two-sided phase fluctuation $\overline{\left<\Delta\phi^2\right>}_{two-sided}$ is smaller than that of one-sided $\overline{\left<\Delta\phi^2\right>}_{one-sided}$ with the definition:
	\begin{equation}
	\label{eq:phiQ}
	\overline{\left\langle\Delta \phi^2\right\rangle}=\frac{1}{T_Q} \int_Q\left< \left( \phi(t)-\hat{\phi}\right)^2\right> d t.
	\end{equation}
	We give the proof in the following. Combining Equation~\ref{eq:phiR} and Equation~\ref{eq:phiQ} the average phase fluctuation of two-sided R-frame is:
	\begin{equation}
	\begin{aligned}
	&\overline{\left<\Delta\phi^2\right>}_{two-sided}
	\\
	=& \frac{1}{T_\mathrm{R}} \int_0^{T_\mathrm{Q}}
	\int_0^{T_\mathrm{R}} \left< \Delta\phi^2(t^\prime+t)\right> \mathrm{d}t^\prime \mathrm{d}t
	\\
	-&\frac{1}{(2T_\mathrm{R})^2}\int_0^{T_\mathrm{Q}}
	\int_0^{T_\mathrm{R}}  \left[  \left< \Delta\phi^2(T_\mathrm{Q}+t^\prime)\right>+\left< \Delta\phi^2(T_\mathrm{Q}+2T_\mathrm{R}-t^\prime)\right>+2\left< \Delta\phi^2(T_\mathrm{R}-t^\prime)
	\right>  \right]  t^\prime \mathrm{d}t^\prime\mathrm{d}t,
	\end{aligned}
	\end{equation}
	while the average phase fluctuation of one-sided R-frame is:
	\begin{equation}
	\begin{aligned}
	&\overline{\left<\Delta\phi^2\right>}_{one-sided}
	\\
	=& \frac{1}{T_\mathrm{R}} \int_0^{T_\mathrm{Q}}
	\int_0^{T_\mathrm{R}} \left< \Delta\phi^2(t^\prime+t)\right> \mathrm{d}t^\prime \mathrm{d}t
	-\frac{1}{(T_\mathrm{R})^2}\int_0^{T_\mathrm{Q}}
	\int_0^{T_\mathrm{R}}  \left< \Delta\phi^2(T_\mathrm{R}-t^\prime)
	\right>   t^\prime \mathrm{d}t^\prime\mathrm{d}t.
	\end{aligned}
	\end{equation}
	So the difference of above two is:
	\begin{equation}
	\begin{aligned}
	&\overline{\left<\Delta\phi^2\right>}_{one-sided}-\overline{\left<\Delta\phi^2\right>}_{two-sided}
	\\
	=&\frac{1}{(2T_\mathrm{R})^2}\int_0^{T_\mathrm{Q}}
	\int_0^{T_\mathrm{R}}  \left[  \left< \Delta\phi^2(T_\mathrm{Q}+t^\prime)\right>+\left< \Delta\phi^2(T_\mathrm{Q}+2T_\mathrm{R}-t^\prime)\right>-2\left< \Delta\phi^2(T_\mathrm{R}-t^\prime)
	\right>  \right]  t^\prime \mathrm{d}t^\prime\mathrm{d}t.
	\end{aligned}
	\end{equation}
	In order to prove the difference is bigger than 0, we only need to prove the core function of integral over Q is bigger than 0:
	\begin{equation}
	\begin{aligned}
	&\int_0^{T_\mathrm{R}}  \left[  \left< \Delta\phi^2(T_\mathrm{Q}+t^\prime)\right>+\left< \Delta\phi^2(T_\mathrm{Q}+2T_\mathrm{R}-t^\prime)\right>-2\left< \Delta\phi^2(T_\mathrm{R}-t^\prime)
	\right>  \right]  t^\prime \mathrm{d}t^\prime
	\\
	>&\int_0^{T_\mathrm{R}}  \left[  \left< \Delta\phi^2(T_\mathrm{Q}+t^\prime)\right>
	-\left< \Delta\phi^2(T_\mathrm{R}-t^\prime)
	\right>  \right]  t^\prime \mathrm{d}t^\prime
	\\
	=&\int_0^{T_\mathrm{R}}  \left< \Delta\phi^2(T_\mathrm{Q}+t^\prime)\right>
	t^\prime \mathrm{d}t^\prime
	-\int_0^{T_\mathrm{R}}  
	\left< \Delta\phi^2(T_\mathrm{R}-t^\prime)
	\right> t^\prime \mathrm{d}t^\prime
	\\
	>&\int_0^{T_\mathrm{R}}  \left< \Delta\phi^2(T_\mathrm{Q}+t^\prime)\right>
	t^\prime \mathrm{d}t^\prime
	-\int_0^{T_\mathrm{R}}  
	\left< \Delta\phi^2(t^\prime)
	\right> t^\prime \mathrm{d}t^\prime
	\\
	=&\int_0^{T_\mathrm{R}} \left[ \left< \Delta\phi^2(T_\mathrm{Q}+t^\prime)\right>
	-\left< \Delta\phi^2(t^\prime)
	\right> \right]t^\prime \mathrm{d}t^\prime
	\\
	>&0,
	\end{aligned}
	\end{equation}
	where the monotonicity of $\left<\Delta\phi^2\right>$ is used in the first and the third inequality, and the rearrangement inequality is used in the second inequality.}

In conclusion, the improved algorithm can recover the spectrum with higher frequency resolution and higher signal-to-noise ratio. As a result, the interference error rate improves from 27.22\% to 3.89\%.	

\section{Laser phase noise characterization}

\begin{figure}[htbp]
	\centering
	\includegraphics[width=\linewidth]{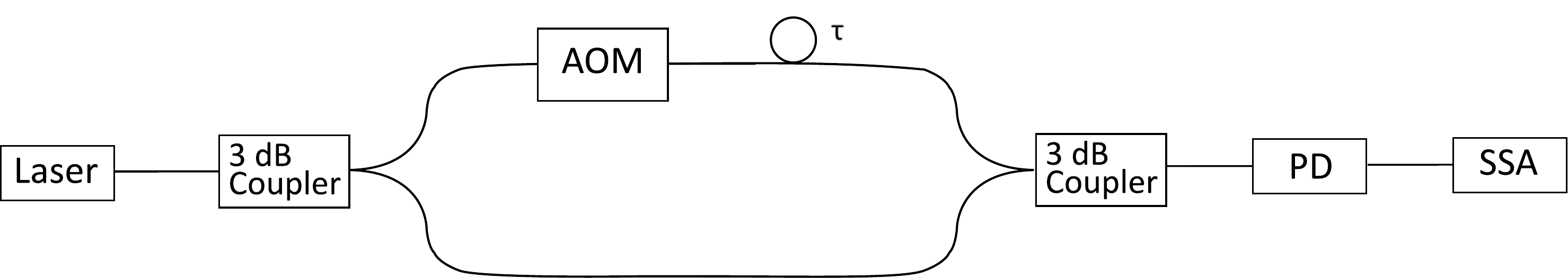}
	\caption{The phase noise measurement setup for lasers. AOM, acoustic optic modulator; PD, photodiode with a trans-impedance amplifier; SSA, signal source analyzer.}
	\label{fig:pn_laser_setup}
\end{figure}

We adopt the delayed self-heterodyne technique~\cite{camatel2008narrow} to measure the laser's phase noise. The setup is shown in Fig.~\ref{fig:pn_laser_setup}. The advantage of this method is that no lengthy fiber (longer than the coherence length of the tested laser) is needed. We use an acoustic optic modulator driven by a 80-MHz sinusoidal signal as the frequency shifter. And the fiber length mismatch is chosen to be 16 m. Therefore a 80-ns delay is introduced, which is enough to ensure that the measured phase noise is mainly contributed by the laser.

\begin{figure}[htbp]
	\centering
	\includegraphics[width=\linewidth]{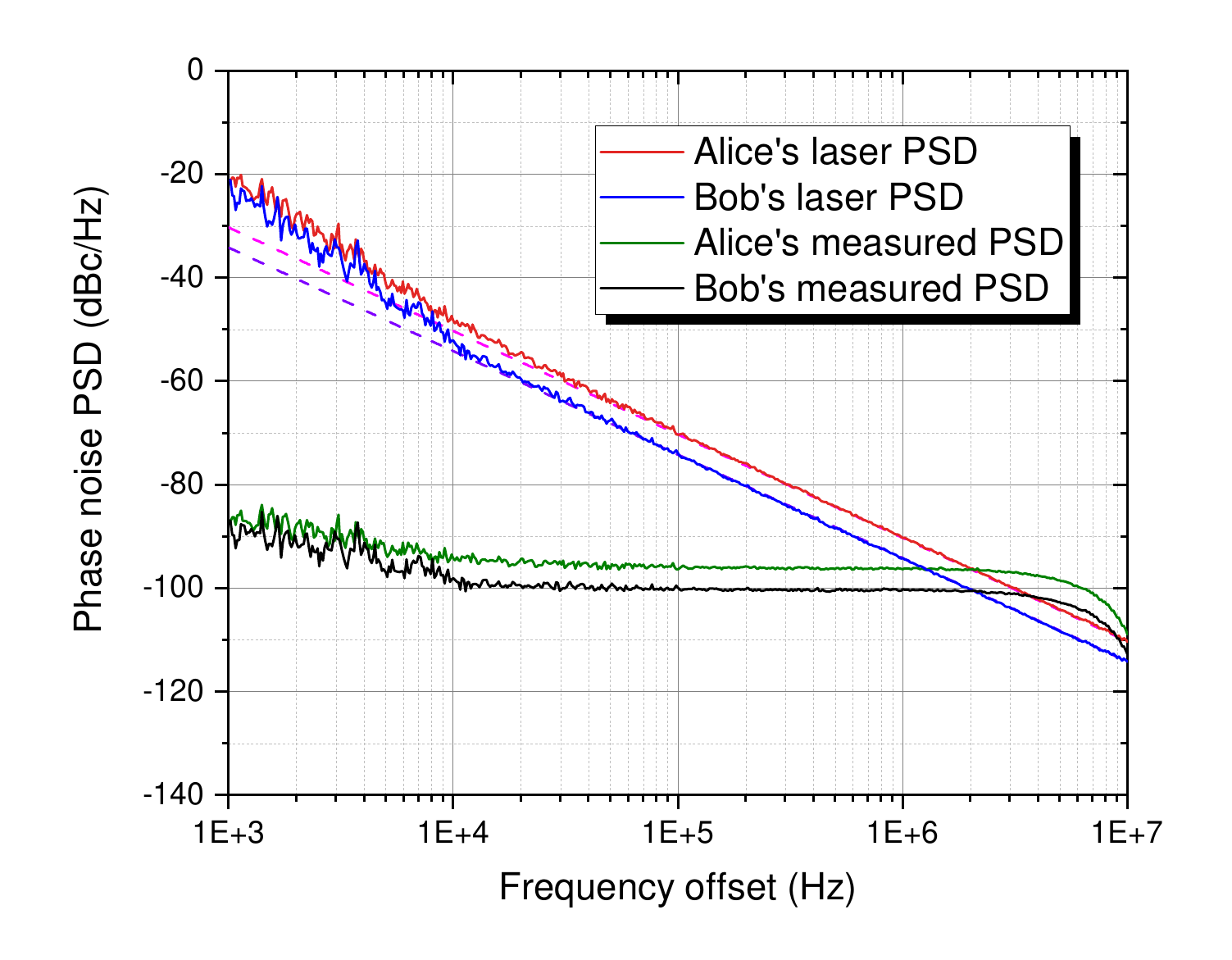}
	\caption{The phase noise PSD for lasers. {The PSD of the lasers $S_{\phi,LD}(f)$ (red and blue lines) are calculated from the measured PSD $S_{\phi}(f)$ (green and black lines) using the setup in Fig.~\ref{fig:pn_laser_setup}.} The white frequency noise PSD is used to fit the data in the frequency range of 100 kHz to 10 MHz.}
	\label{fig:pn_laser_result}
\end{figure}

The measurement is performed on a Keysight E5052B signal source analyzer. The numbers of correlation and trace average are both set to 10. The measured results are shown in Fig.~\ref{fig:pn_laser_result}. The phase noise PSD of the laser $ S_{\phi, \mathrm{LD}}(f) $ is related to measured PSD by the following equation:
\begin{equation}
S_{\phi, \mathrm{LD}}(f)=S_{\phi}(f) \cdot \frac{1}{4 \sin ^2(\pi f \tau)}
\end{equation}
The PSDs of the lasers are also plotted in Fig.~\ref{fig:pn_laser_result}. The PSD of the lasers have a -20 dB/decade slope, indicating that the white frequency noise is the noise source. Therefore, the linewidth of the laser can be simply calculated as $ \delta \nu=\pi h_{\mathrm{w}} $. $ h_{\mathrm{w}} $ is the amplitude of the white frequency noise. The linewidth of Alice's and Bob's lasers are 5.9 and 2.4 kHz respectively.

\section{Fiber phase noise characterization}
\begin{figure}[htbp]
	\centering
	\includegraphics[width=\linewidth]{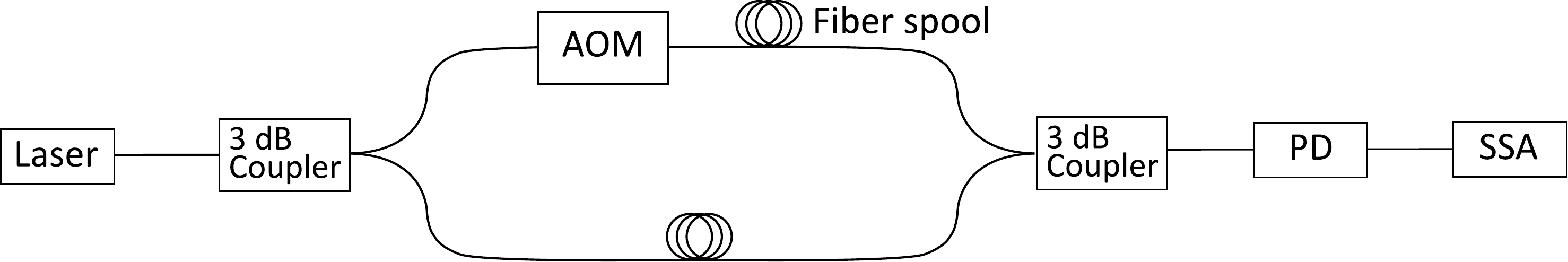}
	\caption{The phase noise measurement setup for fiber spools. The setup is similar to the one for characterizing laser source, except that lengthy fiber spools are inserted in the Mach-Zehnder interferometer {and a lower noise laser (linewidth $ < $100 Hz) is used.}}
	\label{fig:pn_fiber_setup}
\end{figure}

\begin{figure}[htbp]
	\centering
	\includegraphics[width=\linewidth]{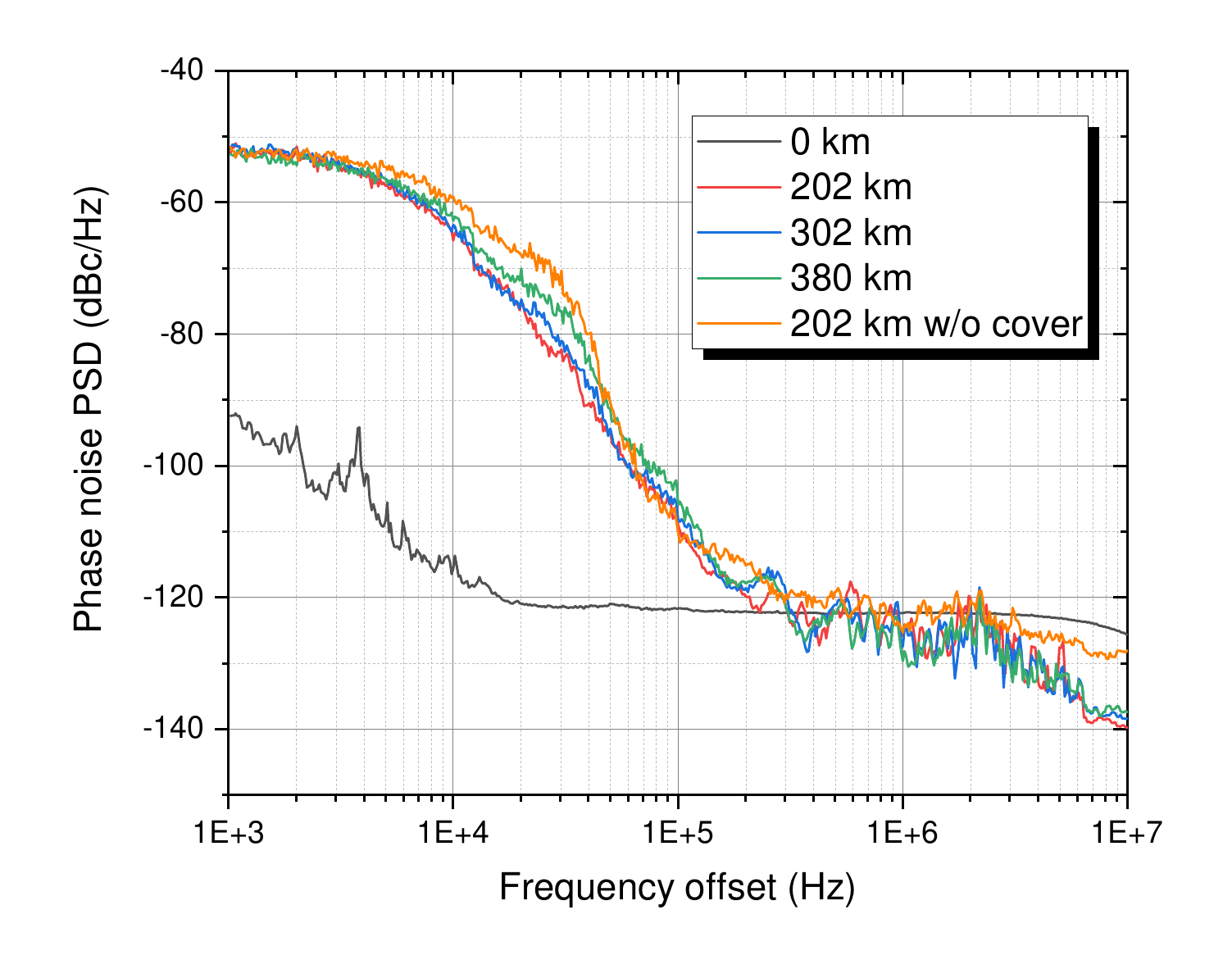}
	\caption{The phase noise results for fiber spools. {The bare fiber spools used in the experiment are covered with foams. One set of spools without any cover are also measured for comparison. The laser source used in the measurement has a linewidth $ < $100 Hz.}}
	\label{fig:pn_fiber_result}
\end{figure}

Also, we characterize the phase noise of the fiber channel using the setup as shown in Fig.~\ref{fig:pn_fiber_setup}. The light source is a fiber laser with a narrow linewidth $ < $100 Hz. This ensure that the measured phase noise is mainly contributed by the fiber spools. {The measured PSDs $ S_{\phi}(f) $ are plotted in Fig.~\ref{fig:pn_fiber_result} and can related to $ \sigma_{L_\mathrm{A}}^2(\tau) $ in Equation~\ref{eq:phase_error} in form: $ 4\pi^2\left(\nu_\mathrm{0A}^2\sigma_{L_\mathrm{A}}^2(\tau)+\nu_\mathrm{0B}^2\sigma_{L_\mathrm{B}}^2(\tau)\right)
	=4\int_{0}^{\infty} S_{\phi}(f)\operatorname{sin}^2(\pi f \tau) \mathrm{d}f $.} The noise below 100 kHz is significantly larger than the one without fiber spools. The noise is mainly contributed by the mechanical vibration of the fibers. As we tap the spool, we observe the noise level below the frequency range of 100 kHz bouncing up and down. {However as the total distance increases from 202 km to 380 km, the increase of the noise at frequency $ > $ 100 kHz is minor. As a result, the interference error rate slightly increases from 0 to 504 km, as we show in the next section. This can be explained by our analytical model. the weighting function $\sin^2(\pi f \tau)$ in the integral acts as a high-pass filter in the frequency domain. In our experiment, $\tau$ is at the microsecond level. The laser has noise level of -90 dBc/Hz @ 1 MHz, while the fiber has noise level of -120 dBc/Hz @ 1 MHz. Therefore, the phase noise contribution from the fiber is minor in our error rate result.} 

\section{Interference error rate characterization}
In order to analyze the sources of the interference error rate reported in the Main Text and validate our theoretical analysis, we perform series of tests under different conditions.

\begin{figure}[htbp]
	\centering
	\includegraphics[width=\linewidth]{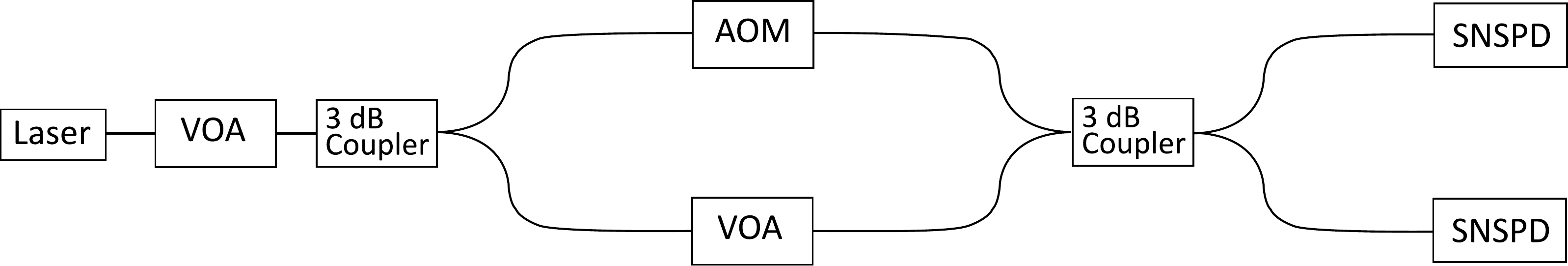}
	\caption{The error rate measurement setup for a single laser. The VOA is used to attenuate the light to single photon level and to balance the loss of the Mach-Zehnder interferometer. VOA, variable optical attenuator; SNSPD, superconducting nanowire single photon detector.}
	\label{fig:onelaser}
\end{figure}

First, we use a self-heterodyne setup as in Fig.~\ref{fig:onelaser} to characterize the error introduced by our fast-Fourier-transform-based algorithm as the phase fluctuations contributed by the source and the fiber channel are negligible in this case. Using the no-phase-locking scheme described in the Main Text, the total error rate is measured to be 1.66\% with a count rate of 24 Mcount/s per detector. 

We use the same setup as in figure 2 in the Main Text to characterize the error rate of two lasers, except that no modulation is applied. In the experiment, $T_\mathrm{R}$ and $T_\mathrm{Q}$ are 4.9152 and 1.6384 $ \upmu $s respectively. As shown in Tab.~\ref{tab:ER}, the count rate is 24 Mcount/s for each detector. The error rate of two lasers increases to 2.55\%. And the fiber channel with different lengths contribute little to the error rate. Therefore, the phase noise of the sources play a bigger role than the fiber channel in our experiment. And to replicate a more practical scenario, we place one of the laser to a different room. The error rate in this case increases slightly.
\begin{table*}[htbp]
	\centering
	\caption{Interference error rate with different configurations.} \label{tab:ER}
	\begin{tabular}{|l|l|}
		\hline
		Configuration     & Interference error rate \\ \hline
		One laser + 0-km fiber     &       1.66 $ \pm $ 0.02\%                  \\ \hline
		Two lasers + 0-km fiber   &      2.55 $ \pm $ 0.06\%                   \\ \hline
		Two lasers + 302-km fiber &         2.64 $ \pm $ 0.17\%                \\ \hline
		Two lasers + 504-km fiber &        2.69 $ \pm $ 0.18\%                 \\ \hline
	\end{tabular}
\end{table*}

\section{Source frequency stability test}
\begin{figure}[htbp]
	\centering
	\includegraphics[width=\linewidth]{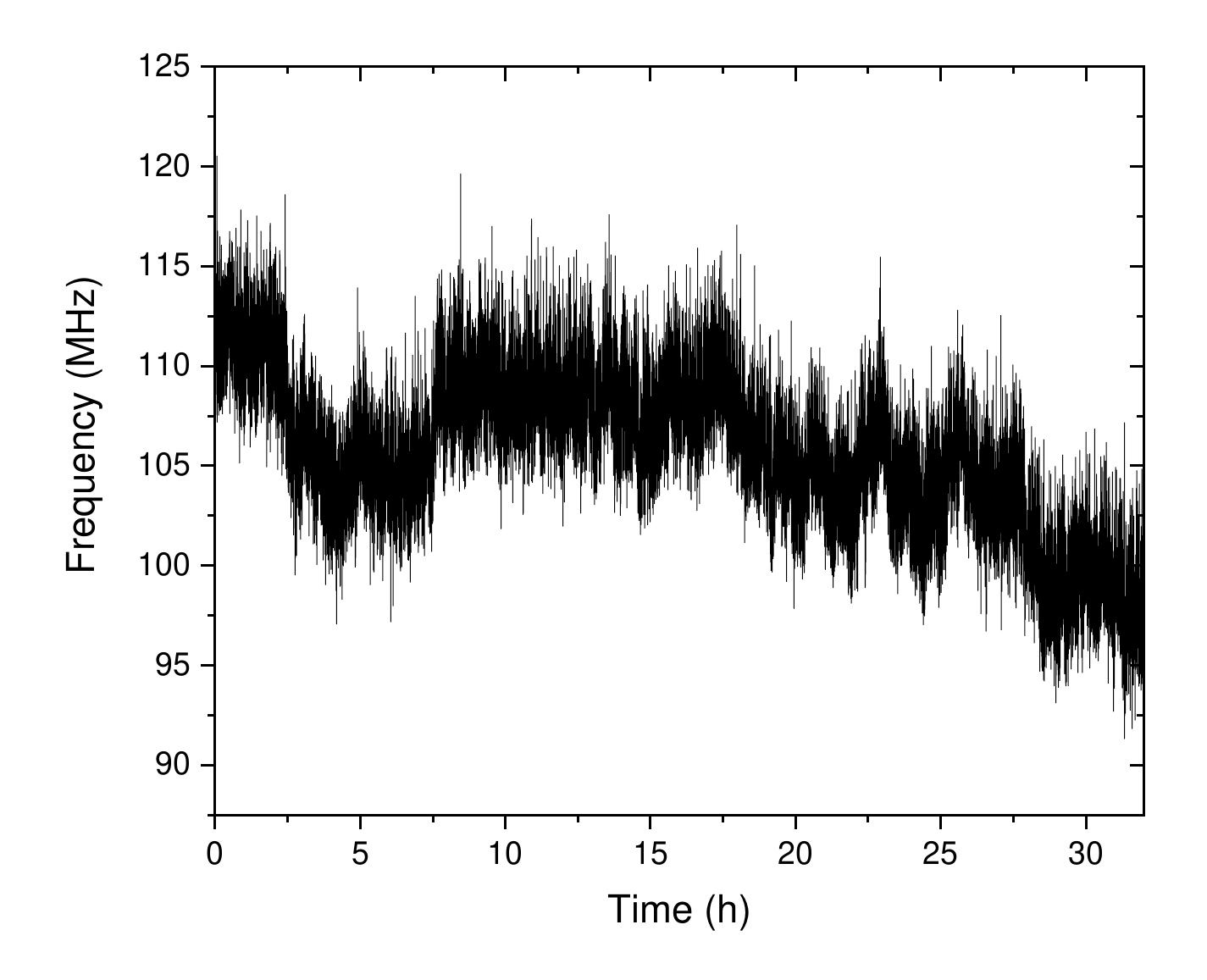}
	\caption{The frequency drift of two unlocked lasers used in our TF-QKD experiment.}
	\label{fig:drift}
\end{figure}

The beat note frequency can be measured by a PD and a frequency counter. The result is shown in Fig.~\ref{fig:drift}. In the estimation of the beat note frequency, we limit the frequency range in 50 to 200 MHz. The frequency is within the range over days. Somehow, if the frequency drifts out of the range, Alice or Bob could abort the QKD session and recalibrate their wavelength by discussing over the public channel.

\section{Three-intensity SNS-TF-QKD with active odd parity pairing protocol}\label{protocol}
The 3-intensity SNS protocol~\cite{wang2018twin,yu2019sending} is implemented by our TF-QKD system, and the actively odd-parity paring (AOPP) method is used to extract the final key~\cite{jiang2021composable}. For the completeness of the paper, we simply review the content as following.

In this protocol, Alice (Bob) has three sources $o,x,y$ ($o^\prime,x^\prime,y^\prime$) with intensities $\mu_o=0,\mu_x,\mu_y$ respectively. Here the sources $o,o^\prime$ are vacuum sources. In each time window, Alice (Bob) randomly decides whether it is a window that emits pulses from source $o$ ($o^\prime$) with probability $p_o$, or a window that emits pulses from source $x$ ($x^\prime$) with probability $p_x$, or a signal window with probability $p_y=1-p_o-p_x$. If it is a signal window, Alice (Bob) randomly chooses the sources $o$ or $y$ ($o^\prime$ or $y^\prime$) with probabilities $1-\epsilon$ and $\epsilon$. For the vacuum pulses in the signal windows, Alice (Bob) denotes them as $0$ ($1$), and for the non-vacuum pulses in the signal windows, Alice (Bob) denotes them as $1$ ($0$).

In the data post-processing, for the time windows that Alice and Bob choose the sources $x$ and $x^\prime$ respectively, Alice and Bob would announce the private phases of the pulse pair and take the following criterion to perform the post-selection:
\begin{equation}\label{cri}
1-\vert \cos(\theta_{A1}-\theta_{B1}-\psi_{AB})\vert\le \lambda,
\end{equation}
where $\theta_{A1}$ and $\theta_{B1}$ are the private phases of Alice's and Bob's pulses respectively; $\psi_{AB}$ can take an arbitrary value, which can be different from time to time as Alice and Bob like;  $\lambda$ is a small positive value. Denote the number of total pulse pairs sent out that satisfactory Eq.~\eqref{cri} by $N_X$ and the number of corresponding error effective events as $m_X$.

We denote the number of pulse pairs of source $\kappa\zeta(\kappa=o,x,y;\zeta=o^\prime,x^\prime,y^\prime)$ sent out in the whole protocol by $N_{\kappa\zeta}$, and the total number of one-detector heralded events of source $\kappa\zeta$ by $n_{\kappa\zeta}$. We define the counting rate of source $\kappa\zeta$ by $S_{\kappa\zeta}=n_{\kappa\zeta}/N_{\kappa\zeta}$, and the corresponding expected value by $\mean{S_{\kappa\zeta}}$. With all those definitions, we have
\begin{equation}
\begin{split}
&N_{oo^\prime}=[p_o^2+2p_op_y(1-\epsilon)]N\\
&N_{ox^\prime}=N_{xo^\prime}=[p_o+p_y(1-\epsilon)]p_xN\\
&N_{oy^\prime}=N_{yo^\prime}=p_op_y\epsilon N
\end{split}
\end{equation}

With the decoy-state analysis, we can get the lower bounds of the expected values of the counting rate of states $\oprod{01}{01}$ and $\oprod{10}{10}$, which are
\begin{align}
\label{s01mean}\mean{\underline{s_{01}}}&= \frac{\mu_{y}^{2}e^{\mu_{x}}\mean{S_{ox^\prime}}-\mu_{x}^{2}e^{\mu_{y}}\mean{S_{oy^\prime}}-(\mu_{y}^{2}-\mu_{x}^{2})\mean{S_{oo^\prime}}}{\mu_{y}\mu_{x}(\mu_{y}-\mu_{x})},\\
\mean{\underline{s_{10}}}&= \frac{\mu_{y}^{2}e^{\mu_{x}}\mean{S_{xo^\prime}}-\mu_{x}^{2}e^{\mu_{y}}\mean{S_{yo^\prime}}-(\mu_{y}^{2}-\mu_{x}^{2})\mean{S_{oo^\prime}}}{\mu_{y}\mu_{x}(\mu_{y}-\mu_{x})}.
\end{align}
Then we can get the lower bound of the expected value of the counting rate of untagged photons
\begin{equation}
\mean{\underline{s_1}}=\frac{1}{2}(\mean{\underline{s_{10}}}+\mean{\underline{s_{01}}}),
\end{equation}
and the lower bound of the expected value of the untagged bits $1$, $\mean{\underline{n_{10}}}$, and untagged bits $0$, $\mean{\underline{n_{01}}}$
\begin{align}
\mean{\underline{n_{10}}}=Np_{y}^2\epsilon(1-\epsilon)\mu_{y}e^{-\mu_{y}}\mean{\underline{s_{10}}},\\
\mean{\underline{n_{01}}}=Np_{y}^2\epsilon(1-\epsilon)\mu_{y}e^{-\mu_{y}}\mean{\underline{s_{01}}}.
\end{align}

And the upper bound of the expected value of the phase-flip error rate satisfy
\begin{equation}\label{e1}
\mean{\overline{e_1^{ph}}}=\frac{\mean{T_{X}}-e^{-\mu_{A1}-\mu_{B1}}\mean{S_{oo^\prime}}/2}{e^{-\mu_{A1}-\mu_{B1}}(\mu_{A1}+\mu_{B1})\mean{\underline{s_1}}},
\end{equation}
where $\mean{T_{X}}$ is the expected value of $T_{X}$, and $T_X=m_X/N_X$.

With values above, we can calculate the lower bound of untagged bits and phase-flip error rate after AOPP, $n_1^\prime$ and $e_{1}^{\prime ph}$ by the method proposed in Refs.~\cite{jiang2021composable}. We have the related formulas of $n_1^\prime$ as follows:
\begin{subequations}
	\begin{align}
	&u=\frac{n_g}{2n_{odd}},\\
	&\underline{n_{10}}=\varphi^L(u\mean{\underline{n_{10}}}),\\
	&\underline{n_{01}}=\varphi^L(u\mean{\underline{n_{01}}}),\\
	&\underline{n_{1}}=\underline{n_{10}}+\underline{n_{01}},\\
	&n_1^r=\varphi^L\left(\frac{\underline{n_{1}}^2}{2un_t}\right),\\
	&n_{01}^\prime=2n_1^r\left(\frac{\underline{n_{01}}}{\underline{n_{1}}}-\sqrt{-\frac{\ln\varepsilon}{2n_1^r}}\right)\\
	&n_{10}^\prime=2n_1^r\left(\frac{\underline{n_{10}}}{\underline{n_{1}}}-\sqrt{-\frac{\ln\varepsilon}{2n_1^r}}\right)\\
	&n_{min}=\min(n_{01}^\prime,n_{10}^\prime),\\
	&n_1^\prime=2\varphi^L\left(n_{min}(1-\frac{n_{min}}{2n_1^r})\right),\\
	\end{align}
\end{subequations}
where $n_t$ is number of raw keys that Alice and Bob get in the experiment; $n_g$ is the number of pair if Alice and Bob perform AOPP to their raw keys; ${n_{odd}}$ is the number of pairs with odd-parity if Bob randomly groups all his raw key bits two by two, and $n_g$ and $n_{odd}$ are observed values; $\epsilon$ is the failure probability of parameter estimation; and $\varphi^U(x),\varphi^L(x)$ are the upper and lower bounds while using Chernoff bound~\cite{chernoff1952measure} to estimate the real values according to the expected values.

And we have the the related formulas of $e_{1}^{\prime ph}$ as follows:
\begin{subequations}
	\begin{align}
	&r=\frac{\underline{n_1}}{\underline{n_1}-2n_1^r}\ln\frac{3(\underline{n_1}-2n_1^r)^2}{\varepsilon},\\
	&e_{\tau}=\frac{\varphi^U(2n_1^r\mean{\overline{e_1^{ph}}})}{2n_1^r-r}\\
	&\bar{M}_s=\varphi^U[(n_1^r-r){e_{\tau}}(1-{e_{\tau}})]+r,\\
	&e_{1}^{\prime ph}=\frac{2\bar{M}_s}{n_1^\prime}
	\end{align}
\end{subequations}

Finally, we can get the key rate $R$ by
\begin{equation}\label{r2}
\begin{split}
R=&\frac{1}{N}\left\{n_1^\prime[1-h(e_{1}^{\prime ph})]-fn_t^\prime h(E^\prime)-2\log_2{\frac{2}{\varepsilon_{cor}}}\right.\\
&\left.-4\log_2{\frac{1}{\sqrt{2}\varepsilon_{PA}\hat{\varepsilon}}}\right\}.
\end{split}
\end{equation}
where $N$ is the total number of pulse pairs sent out in the experiment; $h(x)=-x\log_2x-(1-x)\log_2(1-x)$ is the Shannon entropy; $f$ is the error correction inefficiency; $n_t^\prime$ is the number of survived bits after AOPP; $E^\prime$ is the bit-flip error rate of the survived bits after AOPP; $\varepsilon_{cor}$ is the failure probability of error correction, $\varepsilon_{PA}$ is the failure probability of privacy amplification, and $\hat{\varepsilon}$ is the coefficient while using the chain rules of smooth min- and max-entropy~\cite{vitanov2013chain}.	

\section{Experimental parameters and data}
\begin{figure}[htbp]
	\centering
	\includegraphics[width=\linewidth]{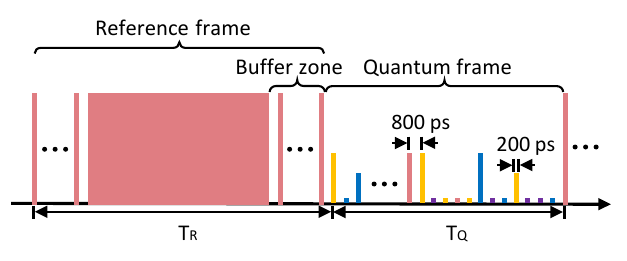}
	\caption{Intensity and phase modulation format. {A buffer zone lasting 20 pulses (16 ns) between the two frames is used to reduce the modulation error from the transition of two frames with different dynamics. The height of the box reflects the intensity modulation and the color reflects the phase modulation.}}
	\label{fig:modulation}
\end{figure}

As Fig.~\ref{fig:modulation} shows, the R-frame is time-multiplexed with the Q-frame, and the ratio of them can be tuned flexibly according to the channel loss. In Tab.~\ref{tab:ratio}, we present the frame configuration for different fiber distances in our experiment. At the long end, the count rate of the reference frame is reduced to lower the dark count noise related to the strong reference pulses. At the short end, the quantum frame duty cycle can be improved to increase the secret key rate (SKR).

\begin{table*}[htbp]
	\centering
	\caption{Frame time and count rate in the reference frame for different fiber distances. $ T_\mathrm{Q} $ ($ T_\mathrm{R} $) is the frame time for the quantum (reference) frame.} \label{tab:ratio}
	\begin{tabular}{|l|l|l|l|}
		\hline
		Fiber distance (km) & $ T_\mathrm{R} $ ($ \upmu $s) & $ T_\mathrm{Q} $ ($ \upmu $s) & Count rate (Mcount/s) \\ \hline
		50                  & 0.4096    & 1.2288    & {48}             \\ \hline
		202                 & 4.9152    & 1.6384    & 24             \\ \hline
		302                 & 4.9152    & 1.6384    & 24             \\ \hline
		380                 & 4.9152    & 1.6384    & 24              \\ \hline
		504                 & 4.9152    & 1.6384    & 8              \\ \hline		
	\end{tabular}
\end{table*}

Tab.~\ref{tab:parameter} summarizes the encoding information of our experiments using 3-intensity SNS protocol. The encoding parameters are optimized to generate higher SKR according to the protocol presented in Section~\ref{protocol}.

The insertion loss of the measurement setup is listed in Tab.~\ref{tab:IL}. The EPC, DWDM and PBS are spliced together to reduce the insertion loss. We use two different SNSPDs for the experiments of 504-km-long fiber and other distances. Their performances are summarized in Tab.~\ref{tab:snspd}. At 504 km, the noise in the key generation basis is mainly from the scattering photons in the fiber induced by the strong light in the R-frame. Therefore, the dense wavelength-division multiplexer is used to reduce the Raman noise. However, double Rayleigh scattering still contributes to the noise count of 160 count/s per detector.

{In Tab.~\ref{tab:data} we report the detailed experimental results for the different fiber distances. Raw detection counts needed to calculate the SKR and and the AOPP results are listed. The counts are labelled in the suffix as `AB$ab$', `A' (`B') and `$a$' (`$b$') denotes the basis and decoy intensity chosen by Alice (Bob) for a certain pulse, where `$y$', `$x$', `$o$' are the signal, decoy and vacuum states respectively.}

\begin{table*}[hbp]
	\centering
	\caption{Parameters for the different fiber lengths, including the decoy intensities and probabilities. $ \mu_y,\ \mu_x,\ \mu_o $ are the signal, decoy and vacuum intensities. $ P_Z $ is the probability of choosing Z basis. $ \epsilon $ is the sending probability when Z basis is chosen. $ P_{\mu_x} $ is the probability of sending in $ \mu_x $ decoy intensity. {$ \gamma $ is the contrast ratio between the intensity of R-frame and $ \mu_y$.}} \label{tab:parameter}	
	\begin{tabular}{|l|l|l|l|l|l|l|l|}
		\hline
		Fiber length (km) & $ \mu_y $ (ph/pulse) & $ \mu_x $ (ph/pulse) & $ \mu_o $ (ph/pulse) & $ P_Z $ & $ \epsilon $ & $ P_{\mu_x} $ & $ \gamma $ (dB) \\ \hline
		50  & 0.53 & 0.028 & 0.00005 & 0.94  & 0.28 & 0.055 & -4\\ \hline
		202 & 0.49 & 0.044 & 0.00005 & 0.878 & 0.28 & 0.115 & 3\\ \hline
		302 & 0.48 & 0.045 & 0.00005 & 0.87  & 0.28 & 0.12  & 12\\ \hline
		380 & 0.45 & 0.050 & 0.00005 & 0.85  & 0.28 & 0.136  & 20\\ \hline
		504 & 0.43 & 0.064 & 0.00005 & 0.77  & 0.28 & 0.2 & 25  \\ \hline		
	\end{tabular}
\end{table*}
\begin{table*}[hbp]
	\caption{Insertion loss of the measurement setup. The results are specified in dB. }
	\label{tab:IL}	
	\centering
	\begin{tabular}{|l|l|l|}
		\hline
		Component & Alice & Bob  \\ \hline
		EPC+DWDM+PBS & 0.87  & 0.92 \\ \hline
		BS-D0        & 3.51  & 3.39 \\ \hline
		BS-D1        & 3.41  & 3.45 \\ \hline
		PC           & 0.21  & 0.12 \\ \hline
	\end{tabular}
\end{table*}   
\begin{table*}[htbp]
	\caption{Performance of two sets of SNSPDs. }
	\label{tab:snspd}	
	\centering
	\begin{tabular}{|l|ll|ll|}
		\hline
		\multirow{2}{*}{Detector} & \multicolumn{2}{l|}{SNSPD\#1}    & \multicolumn{2}{l|}{SNSPD\#2}    \\ \cline{2-5} 
		& \multicolumn{1}{l|}{D$ _0 $} & D$ _1 $ & \multicolumn{1}{l|}{D$ _0 $} & D$ _1 $ \\ \hline
		Efficiency                & \multicolumn{1}{l|}{74\%} & 65\% & \multicolumn{1}{l|}{70\%}     &   72\%   \\ \hline
		Dark count (count/s)      & \multicolumn{1}{l|}{7}    & 6    & \multicolumn{1}{l|}{7}     &   5  \\ \hline
		Dynamic range (dB)        & \multicolumn{1}{l|}{44}   & 48   & \multicolumn{1}{l|}{58}     &  60  \\ \hline
	\end{tabular}
\end{table*}

\begin{table*}[htbp]
	\scriptsize
	\caption{Experimental results and SKR results of finite-size SNS protocol with AOPP. }
	\label{tab:data}
	\centering
	\begin{tabular}{c|cccc|c}
		\toprule
		\toprule
		Detector & \multicolumn{4}{c|}{SNSPD\#1} & SNSPD\#2\\
		\hline
		Distance (km)       & 50        & 202         & 302         & 380         & 504\\
		Loss (dB)           & 9.6       & 38.4        & 56.8        & 72.1        & 96.8\\
		R (bit per pulse)   & $1.35\times10^{-3}$       & $2.15\times10^{-5}$        & $1.18\times10^{-6}$        & $1.25\times10^{-7}$        & $6.56\times10^{-9}$\\
		K (bit per second)  & $1.27\times10^{6}$       & $6.72\times10^{3}$        & $3.69\times10^{2}$        & $3.92\times10^{1}$        & $2.05$\\
		\hline
		$\mathrm{N}_{\mathrm{send}}$         & $10^{10}$  & $10^{10}$  & $10^{11}$  & $10^{12}$ & $10^{13}$\\
		Detected-ZZ$yy$     & 91070688   & 3753305    & 4348062    & 6061368     & 2848762 \\
		Detected-ZZ$yo$     & 125590483  & 4799465    & 5625148    & 7930021     & 3744013 \\
		Detected-ZZ$oy$     & 119406794  & 4855778    & 5423145    & 7773735     & 3470979 \\
		Detected-ZZ$oo$     & 328365     & 13077      & 35858      & 123832      & 68929   \\
		Detected-ZX$yx$     & 10890616   & 950371     & 1162482    & 1957861     & 1529226 \\
		Detected-ZX$yo$     & 935187     & 55645      & 89416      & 613864      & 235751  \\
		Detected-ZX$ox$     & 1510180    & 215377     & 248739     & 194847      & 515998  \\
		Detected-ZX$oo$     & 2457       & 153        & 563        & 5898        & 4721    \\
		Detected-XZ$xy$     & 10228682   & 955528     & 1144738    & 1954830     & 1553976 \\
		Detected-XZ$xo$     & 1543109    & 202480     & 261767     & 199155      & 496252  \\
		Detected-XZ$oy$     & 889432     & 55600      & 86138      & 533252      & 216947  \\
		Detected-XZ$oo$     & 4981       & 303        & 1045       & 3235        & 6753    \\
		Detected-XX$xx$     & 239419     & 72844      & 84062      & 251290      & 337914  \\
		Detected-XX$xo$     & 11791      & 2406       & 4066       & 15654       & 31099   \\
		Detected-XX$ox$     & 11074      & 2572       & 3985       & 13337       & 32511   \\
		Detected-XX$oo$     & 16         & 2          & 9          & 158         & 178  \\
		\hline
		Detected XX$xx$ matching & 30014        & 9055          & 10447          & 31442          & 42569 \\
		Correct XX$xx$ matching & 28560        & 8407          & 9549          & 28859           & 40621 \\
		Sifted key bits in Z-basis before AOPP & 336396330        & 13421625          & 15432213          & 21888956   & 10132683\\
		$\mathrm{QBER}_{\mathrm{ZZ}}$ before AOPP & 27.17$\%$        & 28.06$\%$          & 28.41$\%$          & 28.26$\%$ & 28.79$\%$\\
		Survived key bits in Z-basis after AOPP & 71375551        & 2707044          & 3137485          & 4522494  & 2111964 \\
		$\mathrm{QBER}_{\mathrm{ZZ}}$ after AOPP & 0.20$\%$        & 0.21$\%$          & 0.52$\%$          & 1.23$\%$  & 1.50$\%$\\
		\bottomrule
		\bottomrule
	\end{tabular}
\end{table*}
\clearpage
\end{document}